\documentclass[aps,twocolumn,nofootinbib,showpacs, prd]{revtex4}
\usepackage{graphicx}
\usepackage{amsfonts}
\usepackage{amssymb}
\usepackage{amsbsy}
\usepackage{amsmath}
\usepackage{latexsym}
\usepackage{natbib}
\usepackage{bm}
\usepackage{color}
\usepackage{float}

\begin{document}
\title{Can a variation of Fine Structure Constant influence the fate of Gravitational Collapse?}
\author{Soumya Chakrabarti\footnote{soumya.chakrabarti@vit.ac.in}}
\affiliation{Vellore Institute of Technology \\ 
Tiruvalam Rd, Katpadi, Vellore, Tamil Nadu 632014 \\
India}


\date{\today}

\begin{abstract}
We show that it is possible to steer clear of a spacetime singularity during gravitational collapse by considering time-variation of a fundamental coupling, in this case, the fine structure constant $\alpha$. We study a spherical distribution of cold dark matter coexisting with other fluid elements, collapsing under its own gravity. The dark matter is written as a scalar field interacting with electrically charged matter. This leads to a time variation of $\alpha$ and as a consequence, a breakdown of local charge conservation within the sphere. The exterior has no such field and therefore, Einstein's GR and standard equivalence principles remain valid. We derive the lowest possible bound on the collapse of this sphere beyond which there is a bounce and dispersal of most of the accummulated matter. We discuss the critical behavior of the system around this point and show that the bound is connected to a length scale of the order of Planck, introduced in the theory for dimensional requirements.
\end{abstract}

\maketitle

\section{Introduction}
Theories of fundamental interactions usually carry a sense of mathematical completeness, indicating (i) underlying principles governing the equations of motion (e.g. a Lagrangian formalism) and/or (ii) degrees of freedom as in symmetries and fundamental couplings. Most of these couplings are pre-assigned parameters with no derivation, taken as `fundamental constants' to assign a characteristic scale of the theory. Therefore, time-variation of any of these constants should lead to modifications in \textit{`what is and isn't natural'}. This idea of variation was first proposed as a hypothesis by Dirac, popular as the `Large Numbers hypothesis’ \cite{dirac}. Since then the scientific community has amassed quite a few attempts to accommodate this hypothesis (see for instance \cite{unz, uzan, chiba}) within theories of fundamental interactions. The most well-known attempt is perhaps the field-theoretic approach allowing variations of gravitational constant \cite{jordan}, namely, the Brans-Dicke (BD) theory \cite{bd}. \\

A variation of the fine structure constant $\alpha = \frac{e^{2}}{\hbar c}$ is more radical in comparison; as it is directly related to the variation of either $e$, permittivity of free space or the speed of light $c$. There are some models on varying speed of light providing possible resolutions to some cosmological issues \cite{moffat, albrecht}, however, inevitably they introduce a breakdown of Lorentz invariance \cite{barrow1, barrow2}. In comparison, accepting a time-variation of $e$ seems more practical as long as we take care of local gauge invariance and causality. A general relativistic framework with this variation allows a breakdown of local charge conservation. This means a modified principle of Equivalence, i.e., standard laws of physics are not the same everywhere. Whether or not $\alpha$ can vary with cosmic time at a Hubble rate has been an interesting question asked since the works of Gamow \cite{gamow}. However, for more accurate analyses one must look into fine structure splittings in radio galaxy emission lines \cite{bahcal}, nuclear mass systematics \cite{dyson, peres} and reactor-generated fission product isotopes \cite{shl}. The present estimates of the variation, $\frac{\Delta \alpha}{\alpha} \sim 10^{-6} H_{0}$, is based on studies of active galactic nucleii (e.g. a BL-Lacertae object \cite{wolfe}), flux ratio analysis of galaxy clusters \cite{bora} and relativistic transitions in molecular absorption lines of Quasar spectra at different redshifts \cite{webb, murphy, chand, nunes, parkinson, doran}. It is a natural intuition to assign this variation to a slowly varying scalar field. This new scalar field is therefore, motivated by requirements of fundmanetal interaction, not adhawk and can serve an interesting purpose. It is now been widey accepted, thanks to observations such as luminosity distance measurement of Supernova \cite{riess, betoul}, that at present the universe is expanding with acceleration. The best possible theory to explain this is to bring in an exotic Dark Energy component, (often written as scalar fields!) which can exert a negative pressure and oppose gravity. Nothing concrete can be produced regarding its' origin or distribution other than the claim that it does not cluster below Hubble scale. It is possible for this Dark energy field to be related to the scalar field(s) responsible for fundamental coupling variation. This also insinuates a feedback mechanism between gravitation and particle physics. A few similar insights can be found in theories of unification, where constants of a theory exist only in higher dimensions while their effective four-dimensional projections can be described as scalar fields having variations in space and time \cite{anton}. 
\\

We focus on a theory proposed by Bekenstein \cite{beken} that combines Maxwell's theory and GR \cite{landau}. It allows $\alpha$ to vary through a scalar field interacting with electrically charged matter. This is a special case of the more recently proposed scalar-matter interacting models, popular as Chameleons \cite{chameleon}. They are remarkably successful in addressing cosmological issues, through standard as well as their extended formalisms \cite{brax, wang, sckd}. Generalization of Bekensteins' varying $\alpha$ theory has also received some success in explaining cosmological issues \cite{barrowalpha, magu}. For instance, it has been proved quite recently that the theory can also be generalized successfully to accommodate a simulteneous variation of gravitational constant and fine structure constant \cite{barrowsand, sc1}. This brings us to an important question : \textit{is it always necessary to introduce an exotic field by hand to account for the dark sectors of our universe?} For example, in the varying $\alpha$ theory, we can imagine that the scalar field responsible for a generalization of Maxwell's electrodynamics also provides an origin of Dark Energy or Dark Matter, depending on its' interactions with other matter components. In other words, different laws of conservation or different sets of equivalence principle might have complied the universe to behave differently in different epochs. For example, a varying $\alpha$ theory can be used to describe cold dark matter with magnetic fields dominating electric fields \cite{barrowalpha}. This leads to a domination of magnetostatic energy driving the $\alpha$-variation during matter-dominated epoch. However, the variation becomes negligible as the universe starts to accelerate and the scalar field interaction changes its profile. It is also possible to link any such variation with Higgs vaccuum expectation value and in turn, to quark mass variations \cite{campbell, calmet}. Phenomenologically, quark-mass variation is constrained using Quasar spectroscopy through a measurement of varying proton-to-electron mass ratio $\mu$, also related to $\alpha$-variation through \cite{flambaum, king, bag, rahmani, dapra, scmu}
\begin{equation}
\frac{\Delta\mu}{\mu} \sim \frac{\Delta\Lambda_{QCD}}{\Lambda_{QCD}} - \frac{\Delta{\nu}}{\nu} 
\sim R\frac{\Delta \alpha}{\alpha}. \label{defR}
\end{equation}
$\nu$ is the vacuum expectation value, directly related to Quark masses and $\Lambda_{QCD}$ is a characteristic scale. $R$, a negative parameter is connected to the high-energy scales related to a theory of unification and must be estimated phenomenologically ($R \sim -50$ according to Avelino et. al. \cite{avelino}). \\

We do not study cosmological solutions or the constraints on variation of standard model parameters in this work. There are quite a few unresolved puzzles in gravitational physics that requires counter-intuitions. We focus on one such particular question, related to the phenomenon of Gravitational Collapse and a subsequent formation of spacetime singularity. Any stellar distribution eventually burns out, i.e., exhausts its nuclear fuel supply. If their energy-momentum distribution is studied according to the field equations of standard GR, one can prove that after the exhaustion they will shrink to a zero proper volume. The formation of zero volume comes alongwith a geodesic incompleteness, divergence of curvature scalars or simply, a singularity. This process is best demonstrated by considering idealized collapsing spherical stars such as massive neutron cores \cite{datt, os}, perfect/imperfect fluids \cite{yodzis, eardley} or scalar fields \cite{christo, gonca, scnb1, scnb2}. In principle, singularities indicate a breakdown of classical principles and a general loss of predictability \cite{joshi}. A singular state may or may not communicate with an observer depending on a number of factors, such as, initial size of the distribution and more importantly, formation of a horizon. This leads to the issue of \textit{Cosmic Censorship} \cite{penrose}, whose resolution has been another popular brain-twister for more than five decades, but unfortunately, it has mostly remained inconclusive. There is one possibility, that a well-motivated extension of GR with a modified stress-energy distribution would generate a dominant repulsive effect during the end-stages of a spherical collapse. This can rule out a formation of singularity classically and drive the star into bounce every time, dispersing away all of the accummulated matter. \\    

We prove that this modified stress-energy distribution need not be exotic. It can be found naturally if we consider a theory of gravity accommodating the variation of fundamental couplings, in this case, the fine structure constant $\alpha$. In reality, we are studying a spherical distribution of ordinary matter, radiation fluid, cold dark matter consisting of electric and magnetic fields and a dark energy fluid. Inside the sphere the magnetostatic energy dominates the other components. The scalar field responsible for $\alpha$ variation leads to modified field equations and a breakdown of local charge conservation. Outside, there is no such field and therefore, Einstein's GR and standard equivalence principles remain valid. We show that for this collapsing system of `interacting scalar field dark matter', there can be no formation of singularity even with spatial homogeneity. There is always a lowest possible bound on the radius of the two-sphere beyond which most of the collapsed matter distribution must bounce and disperse. This cutoff scale is connected to a length scale of the order of Planck which is introduced in the theory for dimensional requirements at the outset. \\

The sections are organizd as follows : section $2$ includes our discussion on the generalized Bekenstein's theory in brief. Section $3$ gives the formalism, equations and solution describing a collapsing sphere of $\alpha$-varying matter. Section $4$ includes a detailed discussion on the matching of this collapsing sphere with a suitable exterior geometry across a boundary hypersurface. A few additional commments and a summary is given in Section $5$.

\section{Generalization of Bekenstein's Theory}
We write the charge of an electron as $e = e_{0}\epsilon (x^\mu)$. $\epsilon$ acts as a dimensionless scalar field while $e_{0}$ provides information related to dimension. In effect, $\epsilon (x^\mu)$ works as a universally evolving field through which any fundamental particle charge can vary, provided that parameters such as $e_{0}$ are assigned to take care of the dimensions. We call this field an $e$-field. We further assume that velocity of light and Planck's constant have no variation and therefore, the resulting $\alpha$-variation leads to a departure from `Maxwellian electrodynamics'. This is better understood from a characteristic $\alpha$-evolution equation which should be derived from an invariant action. The evolution equation must be second-order and hyperbolic in nature to avoid issues such as non-causality or runaway solutions. We review the mathematical formulation following Bekenstein's original work on the dynamics of a charged particle in flat spacetime \cite{beken, barrowalpha}. \medskip

With a rest mass $m$ and charge $e_{0}\epsilon$, a particle has a Lorentz-invariant Lagrangian
\begin{equation}\label{actionreview}
L = -mc(-u^{\mu}u_{\mu})^{\frac{1}{2}} + \frac{e_{0}\epsilon}{c}u^{\mu}A_{\mu}.
\end{equation}
We use $\tau$ to express proper time and define $u^{\mu} = \frac{dx^{\mu}}{d\tau}$ as four-velocity. The vector potential term is minimally coupled, making the Lagrangian invariant under a gauge transformation 
\begin{equation}
\epsilon A_\mu = \epsilon A_\mu + \chi_{,\mu}.
\end{equation}

From the action we can write the Lagrange equation as
\begin{equation}\label{lagrangereview}
\frac{d}{d\tau}\left[m u_{\mu} + \frac{e_{0}}{c}\epsilon A_{\mu} \right] = -m_{,\mu}c^{2} + \frac{e_{0}}{c}(\epsilon A_{\nu})_{,\mu}u^{\nu},
\end{equation}
where the normalization $u_{\mu}u^{\mu} = -c^{2}$ is used. Eq. (\ref{lagrangereview}) can be simplified into
\begin{equation}\label{lagrangereview2}
\frac{d(m u_{\mu})}{d\tau} = -m_{,\mu}c^{2} + \frac{e_{0}}{c} \left[ (\epsilon A_{\nu})_{,\mu} - A_{\mu})_{,\nu} \right] u^{\nu}.
\end{equation}
We identify the term $m_{,\mu}c^{2}$ on the RHS as an \textit{anomalous force} term. The Lorentz force term on the RHS provides a gauge-invariant electromagnetic field and lagrangian, written as
\begin{eqnarray}
&& F_{\mu\nu} = \frac{\left\lbrace (\epsilon A_\nu)_{,\mu} - (\epsilon A_\mu)_{,\nu}\right\rbrace}{\epsilon}, \\&&
{\cal L}_{em} = -F^{\mu\nu}F_{\mu\nu}/4.
\end{eqnarray}

A separate lagrangian to govern $\epsilon$-evolution was introduced by Bekenstein, 
\begin{equation}
{\cal L}_\epsilon = -\frac{1}{2}\omega \frac{(\epsilon_{,\mu }\epsilon^{,\mu})}{{\epsilon^2}}.
\end{equation}

$\omega = \frac{\hbar c}{l^2}$ is a parameter required to satisfy dimensional consistency. $l$ is treated as a length scale of the theory which defines a lower limit below which the electric field for a point charge can not be Coulombic. As a consequence, the corresponding energy scale $\frac{\hbar c}{l}$ is also constrained. We use a transformed gauge to generalize this setup 
\begin{eqnarray}
&& a_\mu = \epsilon A_\mu, \\&&
f_{\mu\nu} = \epsilon F_{\mu\nu} = \partial_\mu a_\nu - \partial_\nu a_\mu.
\end{eqnarray}

We replace $\epsilon$ by a $\psi$-field, where $\psi = ln\epsilon$. The combined action can be written as
\begin{eqnarray}\label{standbsbn}
&& S = \int d^4x \sqrt{-g}\Big({\cal L}_g+{\cal L}_{mat}+{\cal L}_\psi +{\cal L}_{em}e^{-2\psi }\Big), \\&&
{\cal L}_\psi = -{\frac{\omega}{2}}\partial_\mu \psi \partial^\mu \psi, \\&&
{\cal L}_{em}=-\frac{1}{4}f_{\mu \nu }f^{\mu \nu}. \\&&
{\cal L}_g = \frac{1}{{16\pi G}}R.
\end{eqnarray}

The action has a similarity with dilatonic scalar field theories \cite{beken2, marciano, damour}, however, this Lagrangian is different because $\psi$ couples only with the electromagnetic part. A lagrangian component for ordinary matter ${\cal L}_{mat}$, is also kept for generality. The usual metric variation and a $\psi$ variation lead to the field equations of the theory
\begin{eqnarray}
&& G_{\mu \nu } = 8\pi G\left( T_{\mu \nu }^{mat} + T_{\mu \nu }^\psi +T_{\mu \nu}^{em}e^{-2\psi }\right), \\&&
\Box \psi = \frac 2\omega e^{-2\psi }{\cal L}_{em}. \label{boxpsi} 
\end{eqnarray}

There are discussions on cosmological solutions of the above set of field equations and possible extensions are proposed in literature. However, formation of collapsed objects in this class of theories has never been addressed before.
 
\section{A Collapsing Spherical Distribution} 
We study the dynamics of an idealized collapsing star in this theory. This idealization means spherical symmetry and spatial homogeneity of the constituent matter components. We do not necessarily mean a spherical region consisting of $\alpha$-variation generating dark matter particles alone and that is perhaps not practical. In reality, any dark matter distribution appears to be in a clustered state around galaxies. They slow down the rotational motion of galaxies by contributing to the inward gravitational pull. This clustering effect competes with the dark energy field/fluid, which plays the role of a driver of cosmic acceleration. It is generally believed that more than $85\%$ of every cluster's mass is contributed in the form of dark matter \cite{kravtsov}. It has also been found that these clumps in distant clusters of galaxies can warp background light from other objects \cite{schmidt}. We believe it is reasonable to assume that an evolving over-dense region of dark matter can co-exist with about $\sim 10\%$ of ordinary matter and dark energy. The cold dark matter is written as a scalar field interacting with electrically charged matter. The dark energy component is written as a pressureless non-interacting field $\rho_{\phi}(t)$. We choose a spatially flat homogeneous metric
\begin{equation}\label{metric}
ds^2 = - dt^2 + a(t)^2(dr^2 + r^2d\Omega^2).
\end{equation}

For this geometry the independent field equations are
\begin{eqnarray}\label{fried1} 
&&\frac{\dot{a}^2}{a^2} = \frac{8\pi }3\Big\lbrace \rho _m\left( 1+\zeta
_m e^{-2\psi}\right) +\rho_{r}e^{-2\psi} \\&&\nonumber
+ \frac{\omega}{2}\dot{\psi}^2 + \rho_{\phi}(t) \Big\rbrace, \\&&
\ddot{\psi} + 3\frac{\dot{a}}{a}\dot{\psi} = -\frac{2}{\omega} e^{-2\psi}\zeta_{m}\rho_{m}.
\label{psiddot}
\end{eqnarray}

The equations are in natural units, i.e., $G = c = 1$. $\zeta_{m} = \frac{L_{em}}{\rho_m}$ is a parameter that determines the percentage of dark matter within the collapsing cluster, compared to the total baryon energy density. Different values to $\zeta_{m}$ can be assigned depending on the comparative strength of electric and magnetic field in the cold dark matter distribution. Cases with $\zeta_{m} < 0$ can describe a cluster of cold dark matter dominated by magnetic coupling, for instance, superconducting cosmic strings \cite{sandvik1}. Cosmologically, this choice is more favored as it can describe the mild variation of $\alpha$ with redshift as observed in molecular absorption spectra of Quasars and at the same time, provide good match with late-time cosmological observations \cite{sc1}. In comparison, $\zeta_{m} > 0$ cases fail to represent the accurate scale of $\alpha$ variation \cite{barrowsand}. At this moment, we do not assign any values to $\zeta_{m}$ and just keep it as a free parameter. The non-interacting ordinary matter and the radiation component satisfy their conservation equations, written as

\begin{eqnarray}\label{dotrhom}
&&\dot{\rho_m}+3\left(\frac{\dot{a}}{a}\right)\rho_m = 0, \\&& 
\dot{\rho_r} + 4\left(\frac{\dot{a}}{a}\right)\rho_r = 2\dot{\psi}\rho_r.  \label{dotrho}
\end{eqnarray}

Eqs. (\ref{fried1}), (\ref{psiddot}), (\ref{dotrhom}) and (\ref{dotrho}) describe the evolution of the collapsing sphere. The fine structure coupling evolves as
\begin{equation}\label{alphadef}
\alpha = \exp (2\psi )e_{0}^2/\hbar c,
\end{equation}
within the sphere and should remain a constant outside. The radius of the two-sphere (coefficient of $d\Omega^2$) is supposed to decrease with time, therefore, $\dot{a} < 0$. We also work under a condition that the $\psi$-field evolution Eq. (\ref{psiddot}) is integrable. This idea is motivated from a mathematical property of a group of second order non-linear differential equations classified as classical anharmonic oscillator equations \cite{duarte, euler1, euler2}. Any Klein-Gordon type differential equation governing a scalar field evolution falls within this class. The analysis involves transforming Eq. (\ref{psiddot}) into an integrable form \cite{harko}. To give an outline we write the general equation as

\begin{equation}\label{anharmonic}
\ddot{\phi}+f_1(t)\dot{\phi}+ f_2(t)\phi+f_3(t)\phi^n = 0.
\end{equation}
$f_1$, $f_2$ and $f_3$ are general functions of any variable, let's say $t$. A pair of transformations convert this equation into an integrable form and are introduced as (provided $n\notin \left\{-3,-1,0,1\right\}$) 
\begin{eqnarray}\label{criterion2}
\Phi\left( T\right) &=&C\phi\left( t\right) f_{3}^{\frac{1}{n+3}}\left( t\right)
e^{\frac{2}{n+3}\int^{t}f_{1}\left( x \right) dx },\\
T\left( \phi,t\right) &=&C^{\frac{1-n}{2}}\int^{t}f_{3}^{\frac{2}{n+3}}\left(
\xi \right) e^{\left( \frac{1-n}{n+3}\right) \int^{\xi }f_{1}\left( x
\right) dx }d\xi .\nonumber\\
\end{eqnarray}%
$C$ is a constant. It can be proved \cite{euler1} that for this transformation to hold true, the coefficients must obey the following condition

\begin{eqnarray}\label{criterion1}\nonumber
&&\frac{1}{(n+3)}\frac{1}{f_{3}(t)}\frac{d^{2}f_{3}}{dt^{2}} - \frac{(n+4)}{\left( n+3\right) ^{2}}\left[ \frac{1}{f_{3}(t)}\frac{df_{3}}{dt}\right] ^{2} \\&&\nonumber
+ \frac{(n-1)}{\left( n+3\right) ^{2}}\left[ \frac{1}{f_{3}(t)}\frac{df_{3}}{dt}\right] f_{1}\left( t\right) + \frac{2}{(n+3)}\frac{df_{1}}{dt} \\&&
+\frac{2\left( n+1\right) }{\left( n+3\right) ^{2}}f_{1}^{2}\left( t\right)=f_{2}(t). 
\end{eqnarray} 

This condition of integrability actually means enforcing an additional symmetry upon the spacetime geometry. Whether or not a scalar field evolution should always be integrable, remains an interesting question. However, it has provided solutions of considerable interest time and again, see for instance recent discussions on scalar fied collapse \cite{scnb1}, self-similarity \cite{scnb2}, cosmology in modified gravity \cite{scjs} and collapse of QCD inspired axions \cite{axion}. We solve Eq. (\ref{criterion1}) directly and use the other field equations to determine the profiles of $\psi$ and the fluid energy density components. We make an approximation $e^{-2\psi} \simeq \gamma(t)\psi^{m} - \gamma_{0}\psi$, where $\gamma(t)$ is a slowly varying function of time and $\gamma_{0}$ is a very small constant ($\gamma_{0} \simeq 10^{-10}$). Therefore scalar field evolution equation becomes

\begin{equation}\label{kgkg}
\ddot{\psi} + 3H\dot{\psi} = -\frac{2\zeta_{m}}{\omega} \gamma \rho_{m} \psi^{m} + \frac{2\zeta_{m}}{\omega} \rho_{m} \gamma_{0} \psi.
\end{equation}  

For $m = -6$, this equation falls within the class of anharmonic oscillator equation with the terms $f_2(t)\phi+f_3(t)\phi^n$ being comparable to to $\frac{2\zeta_{m}}{\omega} \rho_{m} \gamma_{0} \psi -\frac{2\zeta_{m}}{\omega} \gamma \psi^{-6}$. Using the value of $m$, we write and simplify Eq. (\ref{criterion1}) as

\begin{equation}
\frac{\ddot{a}}{a} + 7\frac{\dot{a}^2}{a^2} + \frac{2\zeta_{m}\rho_{0}\gamma_{0}}{\omega}\frac{1}{a^3} = 0.
\end{equation}

A first integral of the above differential equation can be derived as
\begin{equation}\label{firstint}
\dot{a} = - \left(a_{0}a^{-14} - \frac{4\zeta_{m}\rho_{0}\gamma_{0}}{13\omega a} \right)^{\frac{1}{2}},
\end{equation} 

where $a_{0}$ is a constant of integration. From Eq. (\ref{firstint}), it is straightforward to infer that for a real $\dot{a}$, 
\begin{equation}
a_{0}a^{-14} \geq \frac{4\zeta_{m}\rho_{0}\gamma_{0}}{13\omega a},
\end{equation}
or
\begin{equation}\label{alimit}
a(t) \geq \left(\frac{4\zeta_{m}\rho_{0}\gamma_{0}}{13\omega a_{0}}\right)^{-\frac{1}{15}}.
\end{equation}

\begin{figure}[t!]
\begin{center}
\includegraphics[angle=0, width=0.40\textwidth]{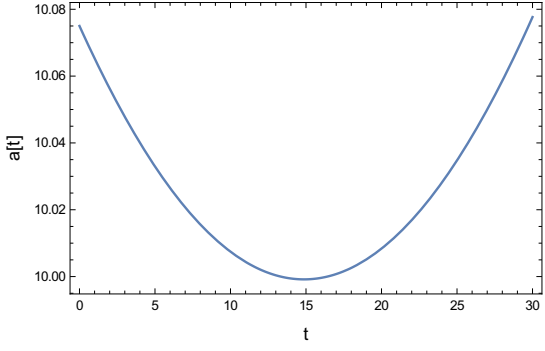}
\caption{Evolution of the radius of two-sphere as a function of time.}
\label{fig1}
\end{center}
\end{figure}

This provides a minimum allowed value of the time evolving factor $a(t)$ until which the stellar body can collapse. After this, the nature of $\dot{a}$ must change and there should be a bounce as we show in Fig. \ref{fig1} through a numerical solution of Eq. (\ref{firstint}). The exact solution for $a(t)$ is found as 

\begin{eqnarray}\nonumber\label{exact}
&& \frac{a\left\lbrace 1 - \frac{a^{13}b_{0}}{a_{0}} \right\rbrace^{\frac{1}{2}}}{8\left\lbrace \frac{a_{0} - a^{13}b_{0}}{a^{14}} \right\rbrace^{\frac{1}{2}}} {_2}F{_1} \left[\frac{1}{2},\frac{8}{13};\frac{21}{13};\frac{a^{13}b_{0}}{a_{0}} \right] = t_{0} - t, \\&&
b_{0} = \frac{4\zeta_{m}\rho_{0}\gamma_{0}}{13\omega}.
\end{eqnarray}

It is not trivial to invert this equation and write $a(t)$ explicitly as a function of time. The parameter $b_{0}$ is crucial as it carries the parameters $\zeta_{m}$ and $\omega$. Just to recall, the choice of $\zeta_{m}$ signifies the nature of cold dark matter within the collapsing sphere. A negative value of $\zeta_{m}$ indicates magnetic interaction dominating the cold dark matter, as in superconducting cosmic strings. On the other hand, $\omega = \frac{\hbar c}{l^2}$ is the parameter introduced to satisfy dimensional consistency of the extended theory. $l$ works as a length scale, a lower limit below which the electric field for a point charge is non-Coulombic. We find from Eq. (\ref{exact}) that for a real evolution, 
\begin{equation}
\frac{\zeta_{m}\rho_{0}\gamma_{0}}{\omega} < 0 ~,~ a_{0} < 0.
\end{equation}

This requirement does not produce any inconsistency in the lower limit of $a(t)$ derived earlier as in Eq. (\ref{alimit}). The first requirement, in particular, means that $\zeta_{m} < 0$ is the only suitable choice for the theory to accommodate a spatially homogeneous \textit{Oppenheimer-Snyder}-type gravitational collapse model, since by definition $\rho_0$, $\omega$ are positive and $\gamma_{0}$ is a pre-defined positive quantity. For all the numerical solutions, we have chosen a particular set of parameters for which $\frac{\zeta_{m}\rho_{0}\gamma_{0}}{\omega} = -0.035$. \\

The fact that a formation of zero proper volume is not possible in this theory even for a spatially homogeneous geometry can also be derived by studying the evolution of kinematic quantities. This is usually done on a slice, or a spacelike hypersurface orthogonal to a congruence of geodesics (see for instance \cite{scsk}). To discuss in brief, we write an induced metric $h_{\alpha\beta}$ to describe this slice
\begin{equation}
h_{\alpha\beta}=g_{\alpha\beta}-u_\alpha u_\beta, \hspace{1.0cm} (\alpha,\beta=0,1,2,3).
\label{eq:transverse_metric}
\end{equation}
The vectors $u^\alpha$ are tangent to the points on each geodesic and timelike. The velocity gradient tensor is defined as
\begin{equation}
B_{\alpha\beta}=\nabla_{\beta} u_\alpha.
\label{eq:vel_grad}
\end{equation}
It is the standard procedure to split $B_{\alpha\beta}$ into three parts. These are `symmetric traceless', antisymmetric and the trace part.
\begin{eqnarray}
&& B_{\alpha\beta}=\frac{1}{3}h_{\alpha\beta}\Theta+\sigma_{\alpha\beta}+\omega_{\alpha\beta}. \\&&\label{eq:bdef}
\Theta = B^\alpha_{\;\alpha}, \\&&
\sigma_{\alpha\beta}=\frac{1}{2}(B_{\alpha\beta}+B_{\beta\alpha})-\frac{1}{3}h_{\alpha\beta}\theta, \\&&
\omega_{\alpha\beta}=\frac{1}{2}(B_{\alpha\beta}-B_{\beta\alpha}).
\end{eqnarray}
$\Theta$ is known as the expansion scalar. $\sigma_{\alpha\beta}$ and $\omega_{\alpha\beta}$ are the shear and rotation tensors, respectively. They satisfy the following relations
\begin{eqnarray}
&& h^{\alpha\beta}\sigma_{\alpha\beta} = 0, ~~ h^{\alpha\beta}\omega_{\alpha\beta} = 0, \\&& 
g^{\alpha\beta}\sigma_{\alpha\beta} = 0, ~~ g^{\alpha\beta}\omega_{\alpha\beta} = 0.
\end{eqnarray}

The spatial tensor $B_{\alpha\beta}$ evolves as
\begin{equation}
u^\gamma \nabla_\gamma B_{\alpha\beta}=-B_{\alpha\gamma}B^\gamma_{\;\beta}+R_{\gamma\beta\alpha\delta}u^\gamma u^\delta.
\label{eq:evolution_eqn}
\end{equation}
The Riemann tensor is written as $R_{\gamma\beta\alpha\delta}$. Trace part of Eq. (\ref{eq:evolution_eqn}) leads to the famous Raychaudhuri Equation \cite{rc}
\begin{eqnarray}
&& \frac{d\Theta}{d\tau} + \frac{1}{3}\Theta^{2} + \sigma^2 - \omega^2 + R_{\alpha\beta}u^\alpha u^\beta = 0, \\&&\label{eq:expansion_eqn}
\sigma^2 = \sigma^{\alpha\beta}\sigma_{\alpha\beta}, \\&&
\omega^2=\omega^{\alpha\beta}\omega_{\alpha\beta}.
\end{eqnarray}

\begin{figure}[t!]
\begin{center}
\includegraphics[angle=0, width=0.40\textwidth]{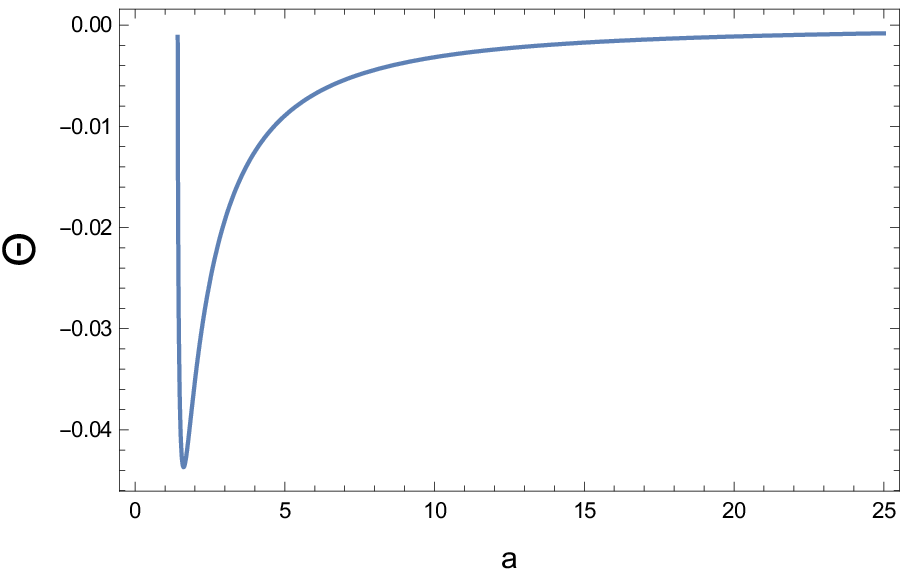}
\includegraphics[angle=0, width=0.40\textwidth]{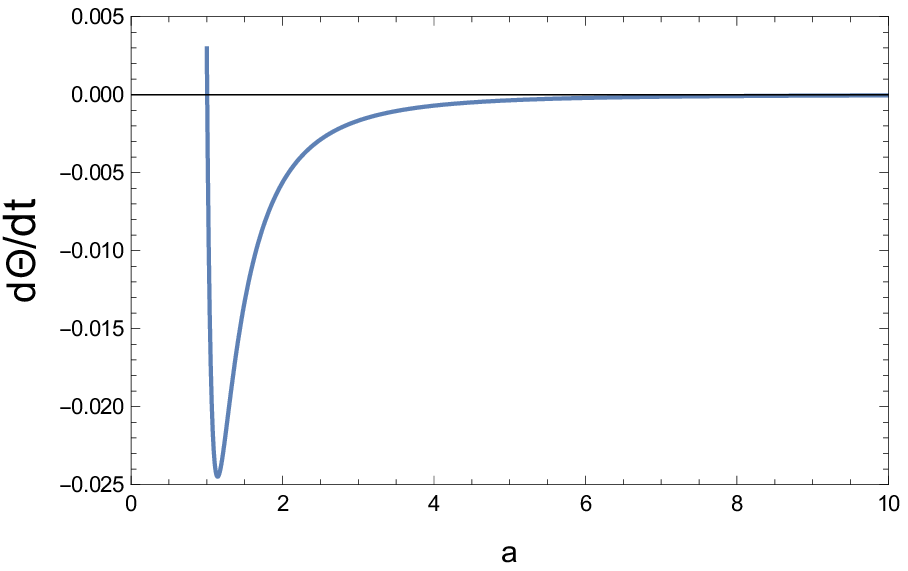}
\caption{Evolution of $\Theta(a)$ as a function of the radius of two-sphere.}
\label{fig11}
\end{center}
\end{figure}

This equation dictates $\Theta$-evolution and its' connection with the spacetime geometry. Unless one uses the Einstein field equations to replace the Ricci tensor $R_{\alpha\beta}$ with energy-momentum tensor, this equation comes purely from Riemannian geometry. $\Theta$, the expansion scalar is very important in particular, as it signifies the distance between two adjacent geodesics on the hypersurface orthogonal. In standard GR, this equation predicts that a family of initially converging geodesics shall focus within a finite time - a result famous as the focusing theorem \cite{rc}. The focusing is understood through the expansion scalar approaching negative infinity, a signature of the formation of a singularity. In any modified theory, this outcome depends on the metric solution, for instance, in our case
\begin{equation}\label{thetageneral1}
\Theta = u^{\alpha}_{;\alpha} = - 3\left(\frac{a_{0}}{a^{16}} - \frac{4\zeta_{m}\rho_{0}\gamma_{0}}{13\omega a^3} \right)^{\frac{1}{2}}.
\end{equation}
For $a_{0} < 0$ and $\frac{\zeta_{m}\rho_{0}\gamma_{0}}{\omega} = -0.035$, the plot of $\Theta$ as a function of $a$ is shown in Fig. \ref{fig11} and it suggests that $\Theta$ never reaches $-\infty$. No real value of expansion scalar is realized near $a(t) \sim 0$. In other words, the collapsing sphere can not shrink too close to zero beyond the minimum cutoff. From the Raychaudhuri Eq. (\ref{eq:expansion_eqn}), we infer that the fate of any geodesic congruence of curves on a collapsing homogeneous sphere ($\sigma^{2} = \omega^{2} = 0$) should be dictated by the signature of $\frac{d\theta}{d\tau}$. If $\frac{d\theta}{d\tau} < 0$, an initially collapsing system will keep on collapsing until $\theta \sim -\infty$. Any transition of the system from collapse into a bounce is understood by a change in signature of $\frac{d\theta}{d\tau}$ into positive which indicates that the geodesics have started to move away from one another. For a family of time-like geodesics, taking the affine parameter $\tau$ as time, we study the evolution of $\frac{d\theta}{d\tau}$ as a function of $a(t)$. The evolution is shown in the lower panel of Fig. \ref{fig11}. It suggests that during the initial phases of the collapse $\frac{d\theta}{d\tau} < 0$ and the rate of collapse increases almost exponentially until the minimum cut-off. After this, $\frac{d\theta}{d\tau}$ starts increasing rapidly and at one point $\frac{d\theta}{d\tau}$ crosses zero to get into positive values. A zero of $\frac{d\theta}{d\tau}$ indicates a critical point $a(t) = a_c$ of the system and can be derived from the equation

\begin{equation}
\frac{d\theta}{d\tau} = \frac{\left\lbrace\frac{a_{0} - a_{c}^{13} \left(\frac{4\zeta_{m}\rho_{0}\gamma_{0}}{13\omega}\right)}{a_{c}^{14}}\right\rbrace^{\frac{1}{2}} \left\lbrace 3 a_{c}^{13} \left(\frac{4\zeta_{m}\rho_{0}\gamma_{0}}{13\omega}\right) - 16a_{0}\right\rbrace}{2a_{c}^{17}\left\lbrace\frac{a_{0} - a_{c}^{13} \left(\frac{4\zeta_{m}\rho_{0}\gamma_{0}}{13\omega}\right)}{a_{c}^{16}}\right\rbrace^{\frac{1}{2}}} = 0.
\end{equation}

\begin{figure}[t!]
\begin{center}
\includegraphics[angle=0, width=0.40\textwidth]{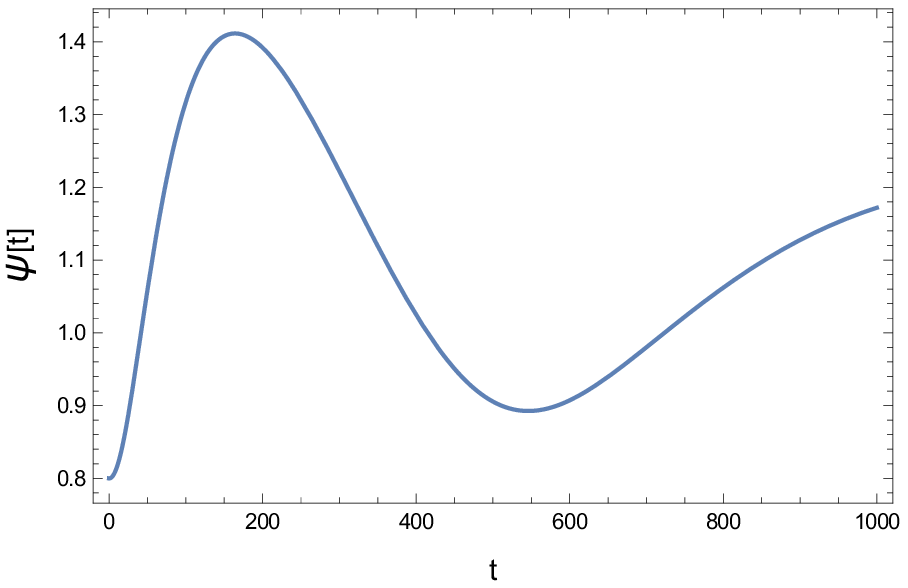}
\includegraphics[angle=0, width=0.40\textwidth]{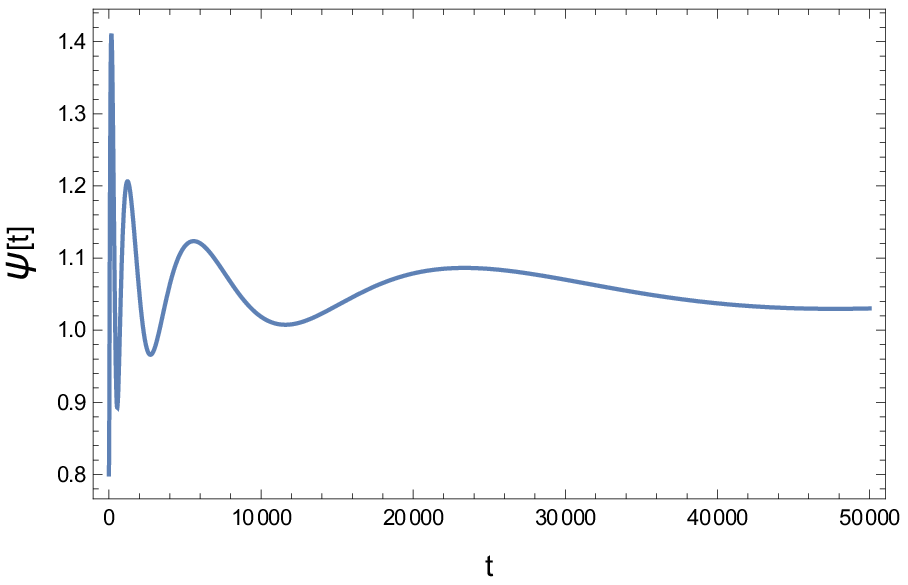}
\caption{Evolution of $\psi(t)$ as a function of time.}
\label{fig2}
\end{center}
\end{figure}

Simplifying this equation (for all $a_{c} > 0$) we find the critical point to be
\begin{equation}
a_{c} = \left(\frac{52 a_{0}\omega}{3\zeta_{m}\rho_{0}\gamma_{0}}\right)^{\frac{1}{13}}.
\end{equation}
Since both $a_{0}$ and $\zeta_{m}$ are negative, the critical point formation is realized at a physical value of $a(t)$. It is  interesting to note that the point of transition of the system depends on four parameters
\begin{enumerate}
\item {$\zeta_{m}$ which signifies the nature of cold dark matter within the collapsing sphere.}
\item {$\omega = \frac{\hbar c}{l^2}$ which is a characteristic length scale of the theory below which the electric field for a point charge is non-Coulombic.}
\item {$\rho_{0}$ which is related to the pressureless dust matter distribution within the collapsing sphere.}
\item {Constant of integration $a_0$ which is most likely directly connected to the initial volume of the collapsing sphere.}
\end{enumerate} 

\begin{figure}[t!]
\begin{center}
\includegraphics[angle=0, width=0.40\textwidth]{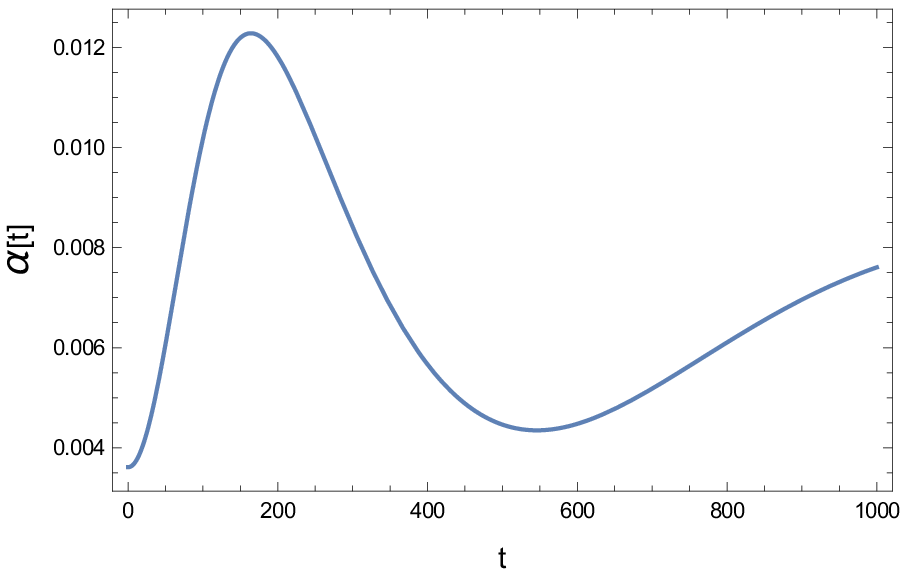}
\includegraphics[angle=0, width=0.40\textwidth]{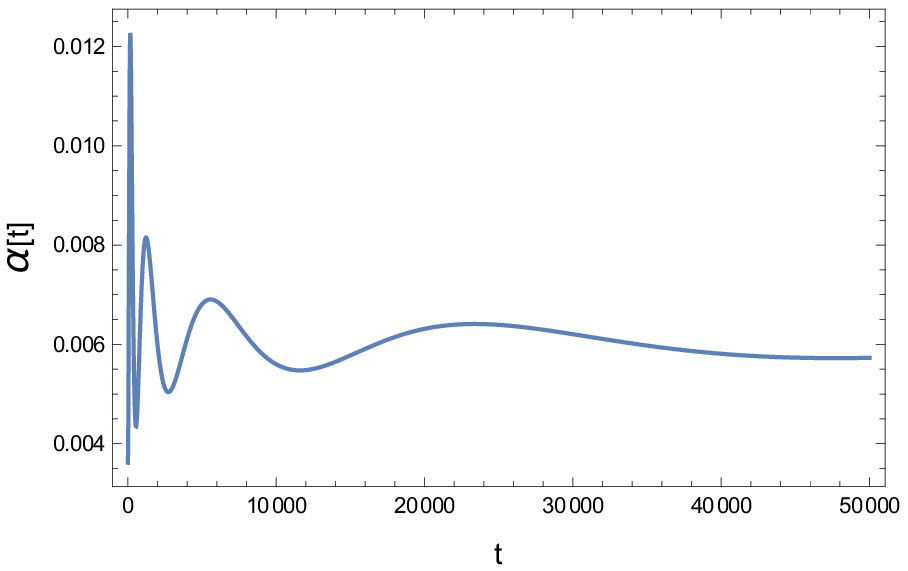}
\caption{Evolution of $\alpha(t)$ as a function of time.}
\label{fig3}
\end{center}
\end{figure}

\begin{figure}[t!]
\begin{center}
\includegraphics[angle=0, width=0.40\textwidth]{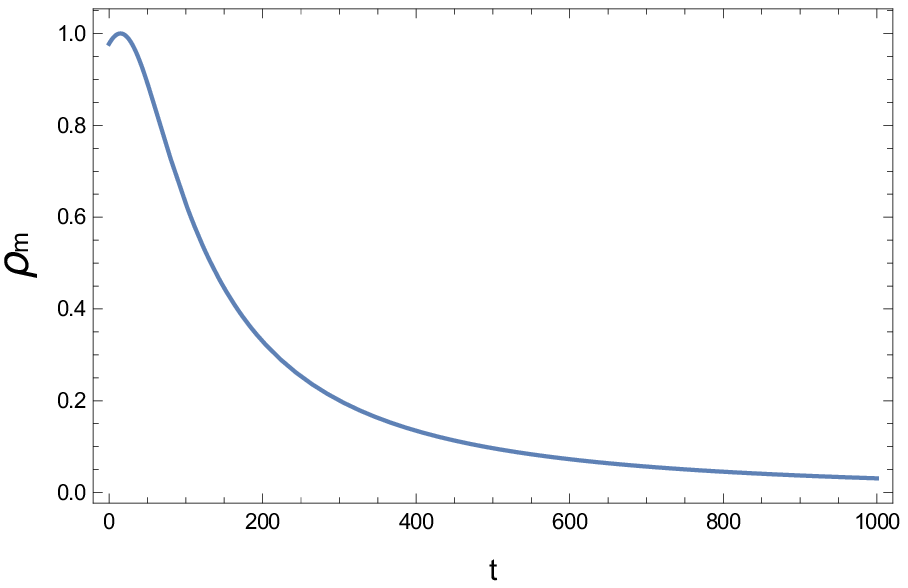}
\includegraphics[angle=0, width=0.40\textwidth]{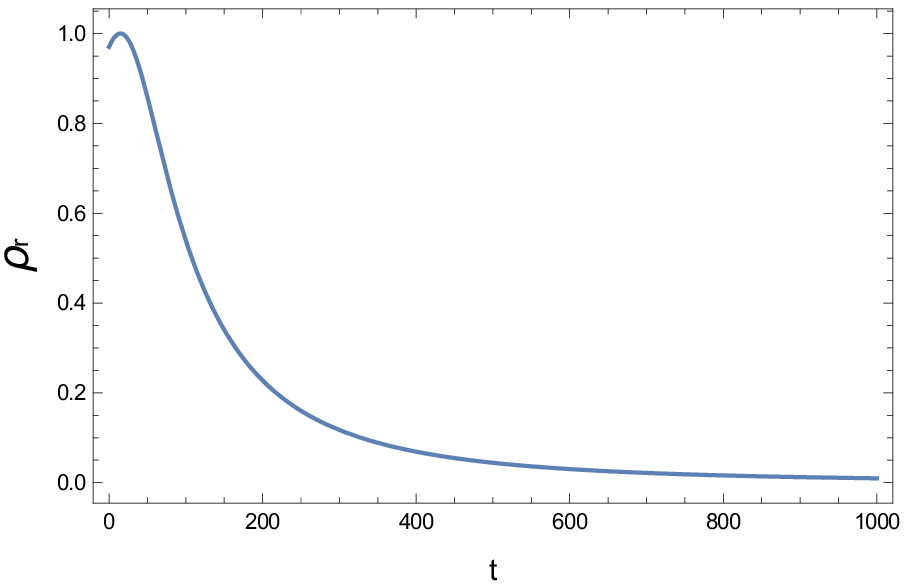}
\caption{Evolution of $\rho_{m}(t)$ and $\rho_{r}(t)$ as a function of time.}
\label{fig4}
\end{center}
\end{figure}

The scalar field $\psi$ must evolve according to the point transformation in Eq. (\ref{criterion2}) and at the same time be consistent with the field Eqs. (\ref{fried1}) and (\ref{psiddot}). Using the numerical solution of Eq. (\ref{firstint}) in Eq. (\ref{psiddot}) we find the consistent evolution of $\psi$. We plot the evolution for different ranges of time in Fig. \ref{fig2}. There is a curious onset of periodicity/oscillation of $\psi$ as the collapsing stellar body starts bouncing. However, The frequency of this oscillation dies down and the scalar field approaches a constant value asymptotically, as almost all the other clustered matter distribution is dispersed away with the bounce. Outside the overdensity, therefore, $\alpha$ or the fine structure constant remains a constant as $\psi$ has no evolution there. However, we must mention that this is a simplified model. Intuitively, for a more general, inhomogeneous collapse $\psi$ should have an evolution $\psi (r,t) \sim \psi_{0}e^{(r_{b}^2 - r^2)f(r,t)}$ where $r_b$ is the boundary of the overdensity. In such a case one should be able to match $\psi$ and it's first derivative across the boundary hypersurface of the collapsing overdensity. The fine structure coupling $\alpha$ evolves exponentially with $\psi$. We plot $\alpha$ in Fig. \ref{fig3}, for different ranges of time. We can see that the periodic behavior of $\psi$ is naturally seen in $\alpha$ evolution as well. Within the stellar body where the matter distribution is dominated by magnetostatic energy, $\alpha$ evolves quite radically and shows oscillations. However, as most of the collapsed matter distribution is dispersed away through a bounce, $\alpha$ asymptotically reaches a constant value $\sim 0.006$, not too far from the value of $\alpha$ we usually assign in classical physics. From Eqs. (\ref{dotrhom}) and (\ref{dotrho}), we also plot the evolution of ordinary matter density and radiation density within the stellar body. Their evolution is shown in Fig. \ref{fig4} and quite naturally shows an initial growth/accummulation during the collapse, before an eventual dispersion to zero value once the bounce takes place. \\

\begin{figure}[t!]
\begin{center}
\includegraphics[angle=0, width=0.40\textwidth]{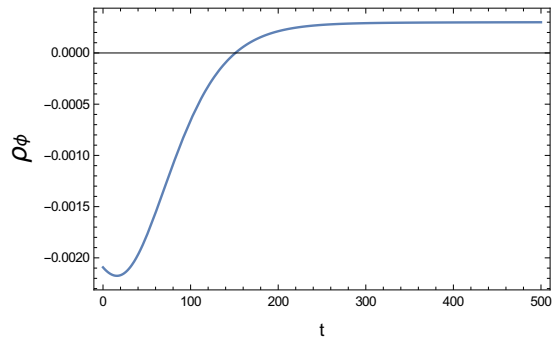}
\caption{Evolution of $\rho_{\phi}(t)$ as a function of time.}
\label{fig5}
\end{center}
\end{figure}
 
In the first field Eq. (\ref{fried1}), we kept one energy density component equivalent to the so-called dark energy distribution and wrote it as a non-interacting fluid/field, $\rho_{\phi}(t)$. Using the solutions of $\psi$ and $a(t)$ we can evaluate the evolution of this field as a function of time. The evolution is shown in Fig. \ref{fig5} and suggests something interesting. During the initial phases of the collapse, this energy density is negative, perhaps playing the role of a negative cosmological constant. However, as the sphere moves into bounce and dispersal phase, $\rho_{\phi}(t)$ goes through a transition from negative into positive domain. Moreover, it becomes a very small positive constant asymptotically. This constant value, at least for this present model is not on the scale of cosmological constant, however, it can suggest of an alternative genesis of the dark energy component. When a massive spherical stellar distribution consisting of (i) cold dark matter driven by magnetostatic energy, (ii) baryonic non-interacting fluid, (iii) radiation and (iv) a dark energy field, collapses under extreme gravity, it will never reach a zero proper volume. It will bounce after a finite time and generate a periodicity of $\alpha$. The periodicity will die down asymptotically until $\alpha$ reaches a constant value. The dark energy density is negative during the implosion, however, as all of the collapsed matter starts dispersing, it evolves into positivity and remains as a remnant with very small positive constant value. Since $\rho_{\phi}(t)$ is not constant for all time, it is better to imagine it interacting non-minimally with geometry or ordinary matter during the initial phases of the collapse and getting decoupled during later phase once the bounce starts. From a simple intuition, the evolution of $\rho_{\phi}(t)$ can be fitted with a functional form such as

\begin{equation}
\rho_{\phi}(t) \simeq \rho_{\phi_0} + \frac{\rho_{\phi_1}}{a^3},
\end{equation}    

where $\rho_{\phi_1}$ is a negative parameter. The nature of this matter component itself can be an interesting topic of discussion as it generates a negative energy density contribution. We comment in passing that the only system known to generate negative energy density is a \textit{quantum inspired Casimir effect}, related to the zero point energy of quantum fields in vacuum. Could such an effect produce the necessary repulsive effects during the critical transition phases of a gravitational collapse and mark the onset of a bounce and dispersal? This is an important question and will be addressed by the author in a separate discussion on quantum corrected gravitational collapse.  \\

We want to mention here that one can easily solve the Klein-Gordon Eq. (\ref{kgkg}) for a different value other than $m = -6$, and the allowed set is quite extensive. However, the equation should fall within the anharmonic oscillator equation class. We give the results for a second example, the $m = -4$ case for which Eq. (\ref{criterion1}) becomes

\begin{equation}
\frac{\ddot{a}}{a} + \frac{2}{13}\frac{\dot{a}^2}{a^2} - \frac{4\zeta_{m}\rho_{0}\gamma_{0}}{13\omega}\frac{1}{a^3} = 0.
\end{equation}

\begin{figure}[t!]
\begin{center}
\includegraphics[angle=0, width=0.40\textwidth]{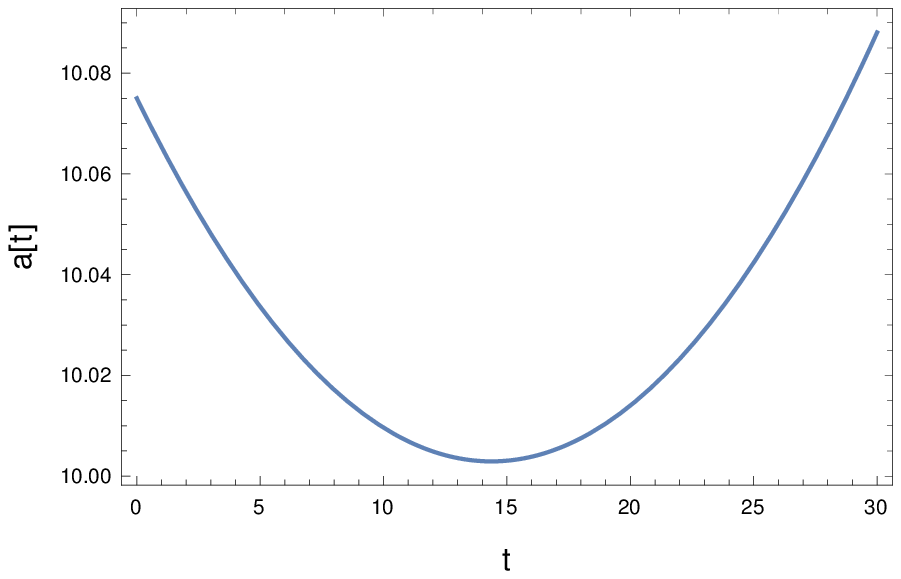}
\includegraphics[angle=0, width=0.40\textwidth]{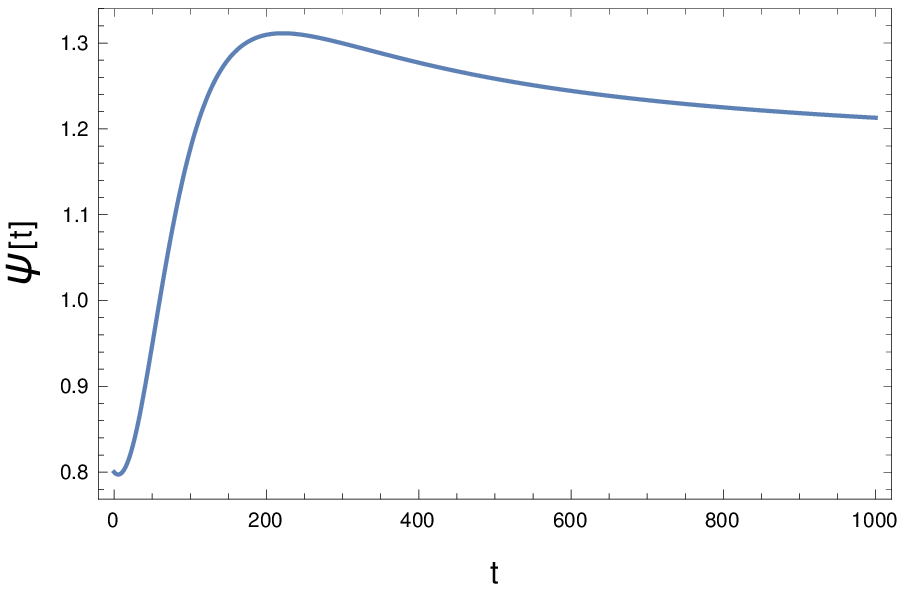}
\includegraphics[angle=0, width=0.40\textwidth]{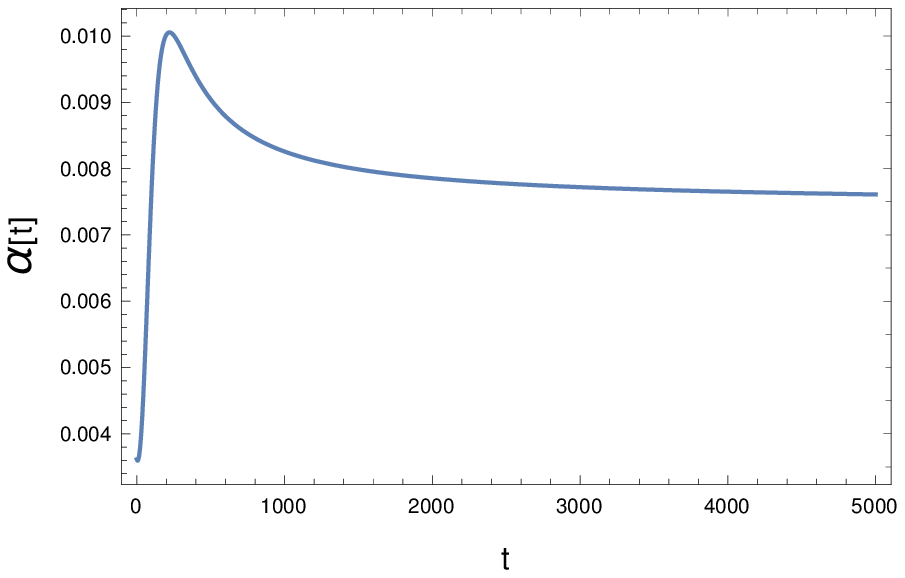}
\caption{Evolution of the radius of two-sphere, $\psi(t)$ and $\alpha(t)$ as a function of time for a second set of initial condition ($m = -4$).}
\label{figic22}
\end{center}
\end{figure}

\begin{figure}[t!]
\begin{center}
\includegraphics[angle=0, width=0.40\textwidth]{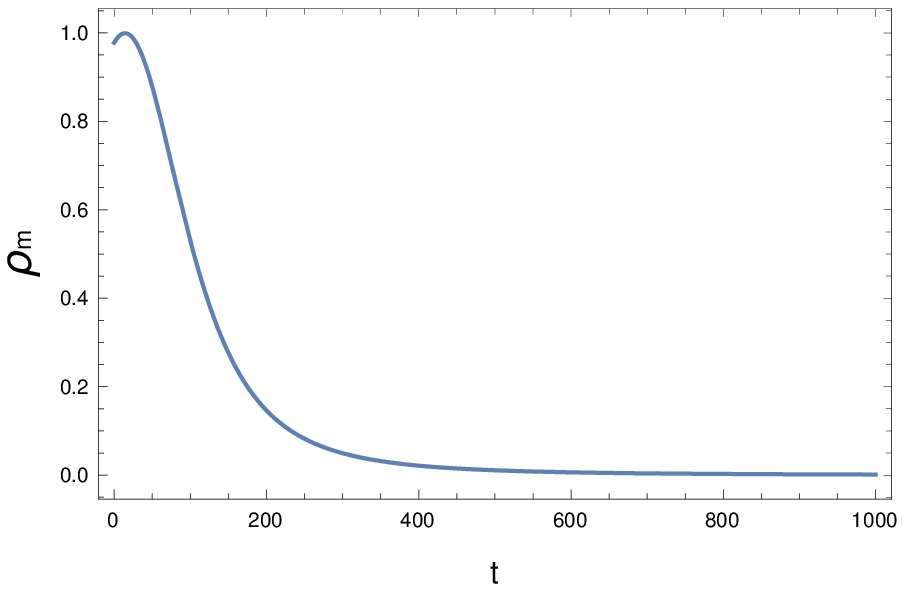}
\includegraphics[angle=0, width=0.40\textwidth]{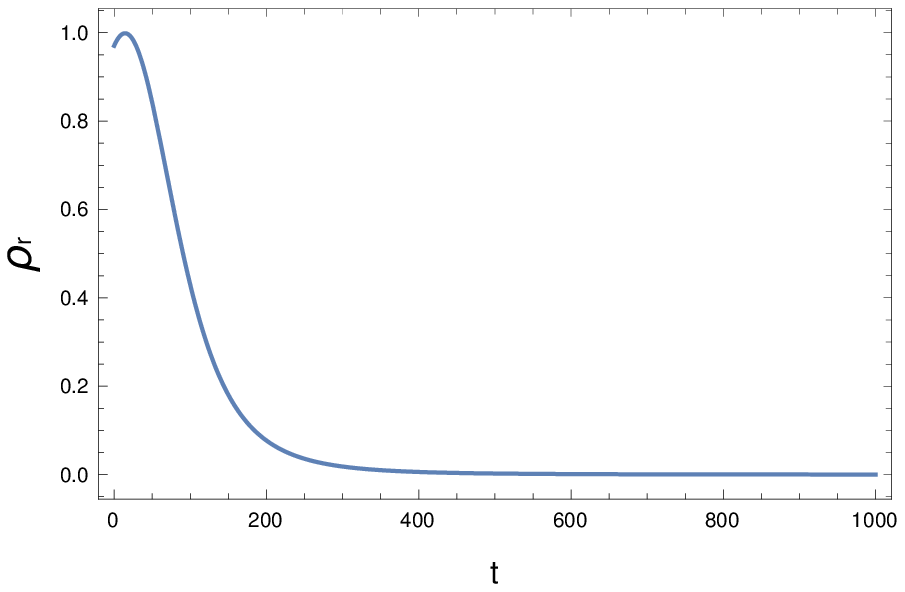}
\caption{Evolution of $\rho_{m}(t)$ and $\rho_{r}(t)$ as a function of time for a second set of initial condition ($m = -4$).}
\label{figic24}
\end{center}
\end{figure}

The radius of the two-sphere for the collapsing star (Fig. \ref{figic22}) describes a similar evolution compared to the $m = -6$ case, i.e., the qualitative non-singular nature of the collapse remains the same. The scalar field $\psi$ evolves as dictated by a consistent solution of Eqs. (\ref{criterion2}), Eq. (\ref{fried1}) and (\ref{psiddot}). A numerical solution of $\psi$ is shown in Fig. \ref{figic22} which suggests that, qualitatively the scalar field approaches a constant value asymptotically as all the clustered matter distribution starts to disperse with the bounce. However, the periodicity/oscillation of $\psi$ within the collapsing stellar body is lost for $m = -4$. The stellar evolution also sees an $\alpha$ evolution asymptotically reaching a constant value $\sim 0.007$ which is again, quite similar to the value of $\alpha$ we usually assign in classical physics (shown in Fig. \ref{figic22}). We also plot the evolution of ordinary matter density and radiation density within the stellar body in Fig. \ref{figic24} and they show an initial growth/accummulation during the collapse, before an eventual dispersion to zero value once the bounce takes place.\\

\begin{figure}[t!]
\begin{center}
\includegraphics[angle=0, width=0.40\textwidth]{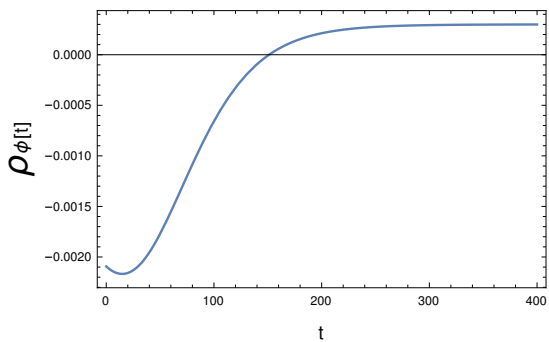}
\caption{Evolution of $\rho_{\phi}(t)$ as a function of time for a second set of initial condition ($m = -4$).}
\label{figic25}
\end{center}
\end{figure}

The dark energy fluid $\rho_{\phi}(t)$ is evaluated numerically and drawn in Fig. \ref{figic25}. Again, it shows a qualitatively same physical behavior compared to the $m = -6$ case, i.e., a negativity during the initial phases of the collapse, a transition into positivity as the sphere moves into bounce and finally, the asymptotic generation of a very small positive constant value.

\section{Matching with an exterior Vaidya Spacetime}
Any collapsing distribution is a system in equilibrium with its exterior and therefore it is crucial to match the two geometries (interior and exterior) across a boundary hypersurface \cite{israel, santos, chan}. Since we have an interacting scalar field inside the collapsing cluster, it is reasonable to assume that the exterior solution can be written using a Vaidya metric. The interior geometry is that of a spatially flat homogeneous metric
\begin{equation}\label{interior}
d{s_-}^2 = - dt^2 + a(t)^2dr^2 + r^2 a(t)^2 d{\Omega}^2,
\end{equation}
while now we define the exterior as
\begin{equation}\label{exterior}
d{s_+}^2 = - \left(1-\frac{2 M(r_v,v)}{r_v} \right)dv^2 - 2dvdr_v + {r_v}^2d{\Omega}^2.
\end{equation}

These two metrics are joined at a boundary hypersurface given by $\Sigma$. The idea is to ensure continuity of the two fundamental forms, metric and extrinsic curvature at this hypersurface. For a general $a(t)$ the metric or the first fundamental form is continuous if
\begin{equation}\label{cond1}
(r_v)_{\Sigma} = r a(t),
\end{equation}
and
\begin{equation}\label{cond2}
\Big(\frac{dv}{dt} \Big)_{\Sigma}=\frac{1}{\sqrt{1-\frac{2M(r_v,v)}{r_v}+\frac{2dr_v}{dv}}}.
\end{equation}

The second fundamental form or the extrinsic curvature is continuous across $\Sigma$ if
\begin{equation}\label{cond3}
\Big(r a(t) \Big)_{\Sigma} = r_v\left(\frac{1-\frac{2M(r_v,v)}{r_v}+\frac{dr_v}{dv}}{\sqrt{1-\frac{2M(r_v,v)}{r_v}+\frac{2dr_v}{dv}}}\right).
\end{equation}

We can combine the above three equations to write
\begin{equation}\label{dvdt2}
\left(\frac{dv}{dt} \right)_{\Sigma} = \frac{3r a(t)^2 - r^2}{3(ra(t)^2 - 2Ma(t))}.
\end{equation}

While Eq. (\ref{cond1}) is popularly accepted as the first matching condition, Eq. (\ref{dvdt2}) serves as the second matching condition. Using Eq. (\ref{cond3}) we can write the Misner-Sharp mass function
\begin{equation}\label{M}
M_{\Sigma} = \frac{1}{4} \Bigg[ra(t) + \frac{r^3}{9 a(t)^3} + \sqrt{\frac{1}{r a(t)} + \frac{r^3}{81 a(t)^9} - \frac{2r}{9 a(t)^5}} \Bigg],
\end{equation}
which provides the total energy confined within the spherical distribution at any value of time or any particular shell of label $r$ \cite{misner}. We also write the rate of change of $M(v,r_v)$ as
\begin{equation}\label{dM}
M{(r_v,v)}_{,r_v}=\frac{M}{r a(t)} - \frac{2r^2}{9 a(t)^{4}},
\end{equation}

using the extrinsic curvature continuity equation. The two equations defining Mass function and it's rate of change are regarded as the third and fourth matching conditions. \\

We also discuss briefly how the exterior geometry should evolve during the collapse of the cluster. If the exterior is written as a generalized Vaidya geometry 
\begin{equation}
ds^{2} = -\left[1 - \frac{2M(u, R)}{R}\right]du^{2} + 2 \epsilon dudR + r^{2} d\Omega^{2},\;\;(\epsilon = \pm 1),
\end{equation}
then $M(u, R)$ gives the energy enclosed within $R$. $\epsilon = \pm 1$ is a parameter that describes different time coordinate choices, namely, Eddington retarded time ($\epsilon = - 1$) and Eddington advanced time $u$ ($\epsilon = 1$). These two choices are two different coordinate representations. For $\epsilon = 1$, $r$ is decreasing along $u = Const$ towards the future. For $\epsilon = - 1$, $r$ is increasing along $u = Const$ towards the future. We write the components of Einstein tensor for this metric as
\begin{eqnarray}
&& G^{0}_{0} = G^{1}_{1} = - \frac{2 M'(u, R)}{R^{2}}, \\&&
G^{1}_{0} = \frac{2 \dot{M}(u, R)}{R^{2}}, \\&&
G^{2}_{2} = G^{3}_{3} = - \frac{ M''(u, R)}{R}.
\end{eqnarray}
The dot is a partial derivative with respect to time coordinate ($\dot{M} \equiv \frac{\partial M}{\partial u}$) and prime is a partial derivative with respect to radial coordinate ($M' \equiv \frac{\partial M}{\partial R}$). We divide the total energy momentum distribution of the exterior in two parts (for more discussions see for instance \cite{hussain1, hussain2, barrabes})

\begin{equation}\label{newvaidyaT}
T_{\mu\nu} = T^{(n)}_{\mu\nu} + T^{(m)}_{\mu\nu}.
\end{equation}
We are using two null vectors $l_{\mu}$ and $n_{\mu}$ to write the Energy Momentum tensor components as
\begin{eqnarray}
&& T^{(n)}_{\mu\nu} = \mu l_{\mu}l_{\nu},\\&&
T^{(m)}_{\mu\nu} = (\rho + P) \left(l_{\mu}n_{\nu} + l_{\nu}n_{\mu}\right) + P g_{\mu\nu}.
\end{eqnarray}

The coefficients (physical quantities) and the null vectors are defined as
\begin{eqnarray}
&& \mu = \frac{2 \epsilon \dot{M}(u, R)}{\kappa R^{2}}, \\&&
\rho = \frac{ 2 M'(u, R)}{\kappa R^{2}}, \\&&
P = - \frac{ M''(u, R)}{\kappa R}.
l_{\mu} = \delta^{0}_{\mu}, \\&& 
n_{\mu} = \frac{1}{2}\left[1 - \frac{2M(u, R)}{R}\right]\delta^{0}_{\mu} - \epsilon \delta^{1}_{\mu}, \\&&
l_{\lambda}l^{\lambda} = n_{\lambda}n^{\lambda} = 0, \\&&
l_{\lambda}n^{\lambda} = - 1.
\end{eqnarray}

This means $T^{(n)}_{\mu\nu}$ is effectively a matter distribution flowing along $u = Constant$ null hypersurface. We can use an orthonormal basis \cite{wangwu} and write the energy momentum tensor of the exterior as  
\begin{eqnarray}
&& E_{0}^{\mu} = \frac{l_{\mu} + n_{\mu}}{\sqrt{2}} ~,~ E_{1}^{\mu} = \frac{l_{\mu} - n_{\mu}}{\sqrt{2}}, \\&&
E_{2}^{\mu} = \frac{1}{r}\delta^{\mu}_{2} ~,~ E_{3}^{\mu} = \frac{1}{r\sin\theta}\delta^{\mu}_{3}.
\end{eqnarray}

Once simplified, this translates into the standard energy momentum tensor representation for the exterior
\begin{equation}
T_{ab} = \left[
\begin{array}{lccl}
\frac{\mu}{2} + \rho& \frac{\mu}{2}& 0 & 0\\
\frac{\mu}{2} & \frac{\mu}{2} - \rho & 0 & 0\\
0 & 0 & P & 0\\
0 & 0 & 0& P\\
\end{array} \right].
\end{equation}

This is a \textit{Type II} fluid energy momentum tensor \cite{hawkingellis}. The Weak and Strong Energy Conditions for this fluid depends upon the mass function of the system through the equations
\begin{equation}
\mu \geq 0, \;\;\; \rho \geq 0, \;\;\; P \geq 0,\; (\mu \neq 0).
\end{equation}
The Dominant Energy Conditions can be written as
\begin{equation}
\mu \geq 0, \;\;\; \rho \geq P \geq 0,\; (\mu \neq 0).
\end{equation}

For a special case $M(u, R) = M(u)$, the energy condition simply becomes a necessary condition
\begin{equation}
\mu \sim -\frac{2\frac{dM(u)}{du}}{\kappa R^{2}} \geq 0.
\end{equation}

During the collapse and dispersal, the matter distribution that remains in the exterior or is dispersed/ejected into the exterior must satisfy this in order to obey the necessary energy conditions.

\section{Conclusion}
Formation and death of a star is a cataclysmic event that can happen in our cosmos, particularly in locally overdense regions. The more technical term to describe this event is gravitational collapse, which, by virtue of GR can be expressed through a set of non-linear differential equations. A solution of these equations provides a picture of what a collapse can produce and more often than not, the outcome is a singularity, a geodesic incompleteness where space-time curvature reaches infinity. If one considers a different theory of gravity, the set of equations are modified and in principle, the solution can portray a different story. We carry a motivation of finding a theory that can produce enough departure such that a formation of spacetime singularity can be avoided. Our proposal is that such a theory can be found by accommodating the idea of Dirac's Large Number Hypothesis within the action of gravity. The hypothesis suggests that it is more natural to allow universal constants to evolve in theories of fundamental forces. We write a generalized theory of scalar-matter interaction where the scalar field can interact only with electrically charged matter and in the process, results in a time evolution of fine structure constant $\alpha$. \\

Ordinarily in GR, an idealized spherical star/stellar distribution, after losing all of its' internal energy, will collapse to a zero proper volume and form a singularity (e.g. massive neutron cores, perfect fluid, scalar fields). This is also realized from the Raychaudhuri equation of congruences for any such stellar distribution. We discuss that in a theory supporting a variation of fine structure constant $\alpha$, formation of a singularity can be avoided. In principle, an $\alpha$ variation can be realized within a cold dark matter distribution dominated by magnetostatic energy. We study an evolving over-dense region of cold dark matter co-existing with ordinary matter and a pressureless non-interacting dark energy field. Inside the distribution, magnetostatic energy dominates the other components and the $\alpha$ variation leads to modified field equations. Outside, there is no such field and therefore, Einstein's GR and standard equivalence principles remain valid. We find an exact solution that describes a spatially homogeneous spherical body collapsing only until a critical radius. Around this critical point the collapsing sphere changes nature and starts bouncing. We show that this lower bound on the radius of the sphere depends on the nature of collapsing matter, in particular, the cold dark matter distribution within the sphere. It also depends on an energy scale of the order of Planck scale (written through the parameter $\omega = \frac{\hbar c}{l^2}$) which was introduced in the theory for dimensional requirements. We believe that this critical point may be connected to quantum gravity constraints, also related to Planck scale. More analysis on this particular question will be included in a subsequent work. \\    

The nature of cold dark matter in a varying $\alpha$ theory is decided through the parameter $\zeta_{m} = \frac{L_{em}}{\rho}$, i.e., the percentage of dark matter present in comparison with the total baryon energy density. Ideally, one can assign different values to $\zeta_{m}$ and that should lead to a different nature of dark matter. For instance, depending on the comparative strength of electric and magnetic field in the cold dark matter distribution, $\zeta_{m}$ can be chosen between $-1$ and $+1$. Cosmologically, $\zeta_{m} < 0$ is more favorable as it can describe the expected mild variation of $\alpha$ as a function of redshift as observed in molecular absorption spectra of Quasars and at the same time provide good match with late-time cosmological observations. Through this work, we also prove that a realistic Oppenheimer-Snyder-type collapsing solution is only possible for $\zeta_{m} < 0$. It means that a cluster of cold dark matter can collapse under gravitational pull and remain non-singular only if it is dominated by magnetic coupling, as in superconducting cosmic strings. \\

The formation of critical point and a transition from collapse into bounce, generates quite a few interesting behavior in the matter constituents of the collapsing cluster. First of all, a non-trivial periodicity in the evolution of $\alpha$ is noted. The periodicity reaches a maximum frequency around the critical point after which the frequency starts to decay and $\alpha$ reaches a constant value asymptotically. This value, although not exactly the value of $\alpha$ we know today, is not too dissimilar either. The distribution of ordinary fluid and radiation density which were clustered during the collapse disperses away to zero alongwith the bounce. The so-called dark energy distribution is written as a non-interacting field. During the initial phases of the collapse, the energy density contribution of this field is negative, perhaps playing the role of a negative cosmological constant. However, as the sphere moves into bounce and dispersal phase, this field goes through a transition and evolves into a very small positive constant. This constant value, is not on the scale of cosmological constant, but it provides a hint that a dark energy field can be generated as a remnant of collapsing overdense regions of clustered matter. Spherical clusters made of cold dark matter and ordinary matter can collapse, bounce and evolve into a constant energy density correction to Einstein's gravity, much like a cosmological constant. Moreover, the origin of a negative energy density can be an interesting topic of research. The only other example of a negative energy density is found in discussions related to zero point energy of quantum fields in vacuum, through a quantum-inspired Casimir effect. It is not a too far-fetched imagination that during gravitational collapse, two adjacent layers of collapsing matter can get arbitarily close to each other, atleast at some value of time during the evolution. Then one might wonder if quantum field theory effects come into play around a scale close to the Planck length and allow the energy density of the inner layer of the collapsing sphere to be negative with respect to the immediate outer layer. Could such an effect produce the necessary repulsive effects during the critical transition phases of a gravitational collapse and mark the onset of a bounce and dispersal? At this moment we keep these comments as possibilities to be explored in the near future.

\end{document}